# Fail-Stop Group Signature Scheme

Yi-Yuan Chiang[1], Wang-Hsin Hsu[*,2], Wen-Yen Lin[2], Jonathan Jen-Rong Chen[1]

[1] *Dept. of Computer Science and Information Engineering, Vanung University, Taoyuan, Taiwan*

[2] *Dept. of Information Management, National Taichung University of Science and Technology, Taichung, Taiwan*

[*]kimble@nutc.edu.tw

*Abstract*

In this paper, we propose a Fail-Stop Group Signature Scheme (FSGSS). FSGSS combines the features of the Group Signature and the Fail-Stop Signature to enhance the security level of the original Group Signature. Assuming that the FSGSS encounters an attack by a hacker armed with a supercomputer, this scheme can prove that the digital signature is indeed forged. Based on the above objectives, this paper proposes three lemmas and proves that they are indeed feasible. First, how does a recipient of a digitally signed document verify the authenticity of the signature? Second, when a digitally signed document is under dispute, how can the group's manager find out the identity of the original group member who signed the document, if necessary for an investigation? Third, how can we prove that the signature is indeed forged following an external attack from a supercomputer? Soon, in a future paper, we will extend this work to make the scheme even more effective. Following an attack, the signature could be proved to be forged without the need to expose the key.

## 1. Introduction

More and more people and organizations are starting to use electronic documents to conduct official government and private business instead of using paper documents. Among other things, this benefits the environment by reducing the use of paper. It also increases the importance of using digital signatures to guarantee the validity, authenticity and integrity of electronic documents and reducing the risk of those documents being forged.

In order to cope with the wide range of potential uses for digital signature technology, the concept of group signing was born. Let's take a real-life example to explain the process of using a Group Signature Scheme: The chief of Taiwan's

Environmental Protection Administration, along with 19 other staff members of the agency, are eligible to digitally sign documents, including those accusing a subordinate unit of breaking the law. In order to safeguard the agency members' neutrality and protect them from interference, each staffer would activate a digital signature key when they release a statement or document representing the administration. The recipient of the document would be able to verify the authenticity of the digital signature, but in the event someone impeached the integrity or validity of a digitally signed document, the identity of the individual who originally signed the document would remain secret.

Companies or other entities cited for violations by the Environmental Protection Administration could file a complaint with the agency to deny that they had violated the law. As part of the review process, it might be necessary to find out the identity of the official who signed the original document making the accusation. Only the manager in the group would have the ability to find out who signed the document. The manager, however, would not have the ability to pretend to be any other group member in order to forge a digital signature.

Chaum et al. [1] concluded that there are three properties of group signatures: (i) Only members of the group can sign messages. (ii) The recipient can verify that it is a valid group signature, but cannot discover which group member made the signature. (iii) If necessary, the signature can be "opened," so that the person who signed the message is revealed.

There are also some favorable features of a group signature scheme that can applied in a range of fields. A digital signature can ensure the validity and authenticity of electronic documents. If the possibility of a document being forged could be reduced, or even if it were proved that the digital signature was indeed forged, the security level of the digital signature could then be enhanced. As for the application of digital signatures, there is another type of fail-stop signature scheme, or "FSS," which can satisfy the above needs.

Research by Nobuaki et al. [2] showed that a FSS scheme has to have at least two security properties: (i) A scheme based on information-theoretic security is secure even against a computationally-unbounded adversary. (ii) If the computational assumption is broken; an honest signer can prove a forgery by virtue of the information-theoretic security.

In this work, a FSGSS is proposed. FSGSS combines all of the functions and features of two schemes: Group Signature (GS) and a Fail-Stop Signature scheme (FSS).

The layout of this paper is as follows: Section 2 is related work. Section 3 is our scheme. Section 4 provides analysis and discussion. And finally, Section 5 concludes the paper and provides a direction for future research.

**2. Related work**

Desmedt proposed the Group-Oriented Cryptosystem concept in 1987. In his research [3], he pointed out that in addition to entities that exist as individuals, there are entities consisting of groups of many individuals, such as hospitals, schools, public institutions and private companies. When these entities issue signed electronic documents, such as certificates, the concept of a digital signature becomes a mechanism to replace signatures on paper documents. The digital signatures could be placed on electronic diplomas, electronic medical records and other official documents released by governmental agencies. The types of documents that carry digital signatures must have the following features: certainty of identity, nonrepudiation and unforgeability.

So the design of the way keys are exchanged and the parameters of the exchanges become particularly important. Although each member in a group has a secret key, the group password must be reused. In other words, individuals in the group cannot exchange their keys during an operation. Instead, they exchange secondary keys derived from their main keys. This ensures the security of the main keys. In addition, members cannot export the group's master key in order to ensure that this key is kept secure. Jonathan et al. [4] developed a new and faster anonymous digital signature system by linking the LUC function with the complexities of discrete logarithm and factorization.

On other hand, a lot of research has focused on the security of ordinary digital signature schemes that rely on a computational assumption. Fail-Stop Signature schemes provide security for a sender against a forger with unlimited computational power by enabling the sender to provide a proof of forgery if it occurs. FSS schemes have been proposed [2][5][6][7]. Kai et al. [8] proposed that a fail-stop scheme could assert a victim's innocence without exposing the $n = p \times q$ secret and would guard against malicious behavior. And more recently, Takashi N. et al. [9] proposed a framework for FSS operating in a multisigner setting and called for a primitive fail-stop multisignature scheme (FSMS). In other words, they combine threshold [10][11] and fail-stop signatures. After the first aggregate signature scheme was proposed, many researchers have tried to propose an efficient aggregate signature scheme.

FSS provides the security for a signer against a computationally unbounded adversary by enabling the signer to provide a proof of forgery. A conventional GS scheme has none of these properties. In this work, we propose a new scheme for integrating FSS and GS in the next section.

### 3. Proposed scheme
**3-1** Initialization

System Center (SC) chooses a primitive $p_0$, satisfies the below equation.

$$p_0 = 4p_1q_1 + 1 \tag{1}$$

Where $p_1, q_1$ are large primitive,

Let $n = p_1q_1$ (2)

SC chooses a number, $g_2 \in Z_n^*$ satisfies

$$g_2^{p_1} \equiv 1 \pmod{p_0} \tag{3}$$

$\{g_2, p_0, n\}$, $\{p_1, q_1\}$ is the public key and secret key of SC, respectively. The detail of initialization process shown in Fig.1

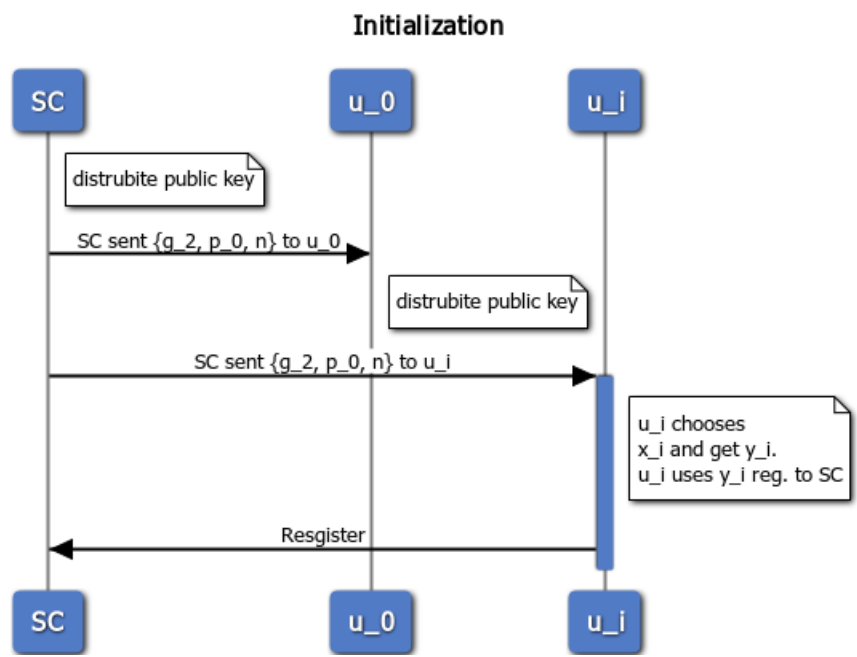

Fig.1    Initialization of GFSS

**3-2 Group and its members**

Without loss of generality, we assume that a group and its members $u_i, 0 \leq i \leq l$, where $u_0$ is manager of group. The member registers to SC as follows individually:

$u_i$ chooses a number $x_i$, and calculates

$$g_2^{x_i} \equiv y_i \pmod{p_0} \tag{4}$$

$u_i$ uses the $y_i$ to register.

**3-3 Parameters exchange**

I. $u_i, 0 \leq i \leq l$, requests a part of parameter from $u_0$. Then $u_0$ chooses a number $k$, and calculates

$$g_2^k \equiv r_1 \pmod{p_0} \tag{5}$$

$$\begin{array}{l} u_0 \to u_i : r_1 \\ \text{(It means } u_0 \text{ send } r_1 \text{ to } u_i\text{)} \end{array} \tag{6}$$

$u_i$ chooses a number $b'$, and calculates

$$g_2^{b'} \equiv b \pmod{p_0} \tag{7}$$

$$r_1^b \equiv r_3 \pmod{p_0} \tag{8}$$

$$r_2 \equiv r_3/b \pmod{n} \tag{9}$$

$$u_i \to u_o : r_2 \tag{10}$$

$u_0$ chooses a number $a$, and satisfying the following equation.

$$a \equiv x_0 r_2 + ks \pmod{n} \tag{11}$$

$u_0 \to u_i : a, s$ (After the above procedure, the manager $u_0$ knew the parameters, $k, a, x_0, r_2, it\ is\ means\ s\ is\ know$) (12)

Note that $y_0, r_1$ is public key of $u_0$, where $g_2^{x_0} \equiv y_0 \pmod{p_0}$, by the equation (4). The detail process of GFSS is shown in Fig.2 3-way handshake for exchange parameters.

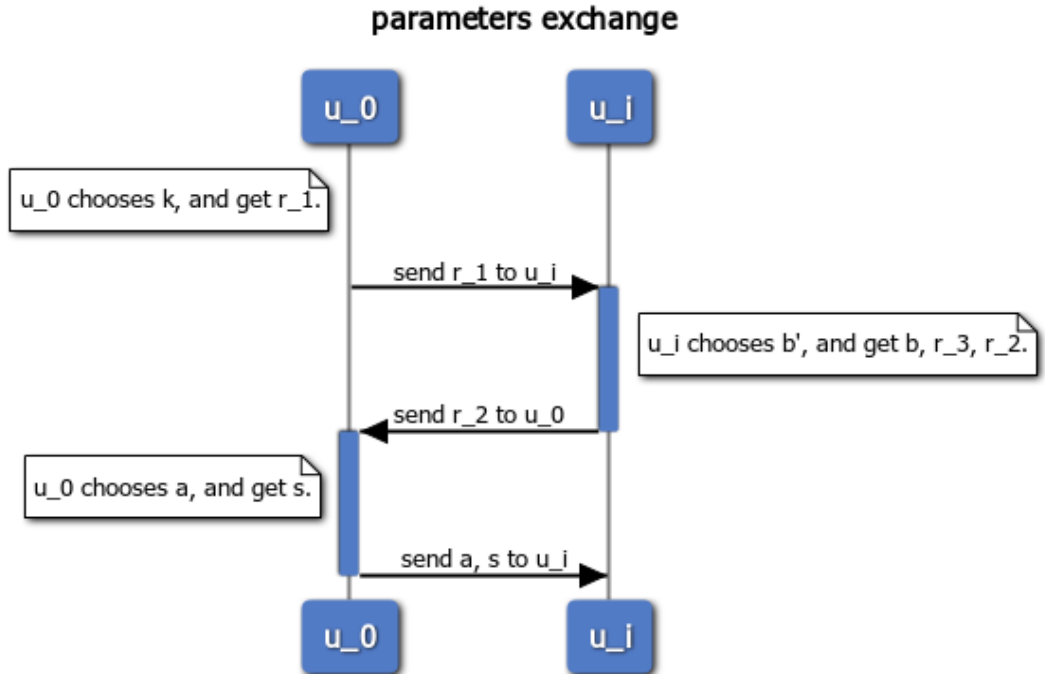

Fig.2  3-way handshake for parameters exchange

**3-4**  $u_i$  Signing a message  $m$

Multiply  $b$  on the 2 sides of the equation (11) and hence,

$$ba \equiv x_0(br_2) + (kb)s \quad (mod\ n) \tag{13}$$

Multiply  $b$  on the 2 sides of the equation (9) and hence,

$$r_3 \equiv br_2 \quad (mod\ n) \tag{14}$$

By equation (13) and (14), we have,

$$ba \equiv x_0 r_3 + (kb)s \quad (mod\ n) \tag{15}$$

chooses 2 numbers  $c, e$  and calculates

$$g_2^c \equiv r_5 \quad (mod\ p_0) \tag{16}$$

$$g_2^e \equiv E \quad (mod\ p_0) \tag{17}$$

Let  $r_4 \equiv r_3 r_5 \quad (mod\ p_0)$ \tag{18}

and  $s_1 \equiv r_5 s \quad (mod\ n)$ \tag{19}

Add  $cs$  on the 2 sides of the equation (15) and hence,

$$ba + cs \equiv x_0 r_3 + (kb)s + cs \tag{20}$$
$$\equiv x_0 r_3 + (kb + c)s \pmod{n}$$

Using $r_5$ in equation (18) and equation (19) to multiply 2 sides of the equation (20) and hence,

$$(ba + cs)r_5 \equiv x_0 \left(\frac{r_4}{r_5}\right) r_5 + (kb + c)\left(\frac{s_1}{r_5}\right) r_5 \tag{21}$$
$$\equiv x_0 r_4 + (kb + c)s_1 \pmod{n}$$

Let $r_6 = (ba + cs)r_5 \pmod{n}$ \qquad (22)

and $m + r_6 = cE + es_2 \pmod{n}$ \qquad (23)

We assume that the recipient of the message is $R$, $u_i$ sends messages to $R$. We note that,

$$u_i \to R : \{m, c, E, r_4, r_6, s_1, s_2\} \tag{24}$$

$R$ calculate the equations as follows.

$$g_2^{r_6} \equiv y_0^{r_4} r_4^{s_1} \pmod{n} \tag{25}$$

$$g_2^{m+r_6} = g_2^{cE} E^{s_2} \pmod{p_0} \tag{26}$$

Recipient accepted this digital signature if both equations above are valid. Otherwise, deny this digital signature.

Table 1. A list of members holding parameters

|  | SC | $u_o$ | $u_i$ | R |  | SC | $u_o$ | $u_i$ | R |
|---|---|---|---|---|---|---|---|---|---|
| $g_2$ | v | v | v | v | $a$ |  | v | v |  |
| $p_o$ | v | v | v | v | $s$ |  | v | v |  |
| $n$ | v | v | v | v | $m$ |  |  | v | v |
| $y_i$ | v | v | v |  | $c$ |  |  | v | v |
| $r_1$ |  | v | v |  | $E$ |  |  | v | v |
| $k$ |  | v |  |  | $r_4$ |  |  | v | v |
| $b'$ |  |  | v |  | $r_6$ |  |  | v | v |
| $b$ |  |  | v |  | $s_1$ |  |  | v | v |
| $r_2$ |  | v | v |  | $s_2$ |  |  | v | v |

## 4 Analysis and discussion

In this section, first we introduce Lemma 1 to check whether a digital signature is valid or not. Lemma 2 verifies whether a digital signature is activated by the group member. Lemma 3 shows the attack method that is mentioned by Willy Susilo et al. [7] will not succeed. There are a lot of parameters after these procedures above. We make a list of members holding parameters as Table 1. In this scheme, members share the partial parameters and keep a few parameter(s). For example, manager $u_0$ only holds the parameter $k$, and a member $u_i$ only holds the parameter $b$. In case someone makes a digital signature of $u_0$ and it is verified but she/he has no idea about $b$, that means someone is a forger.

**Lemma 1** Assumes that $u_0, u_i$ are honest, if both equations (25) and (26) are valid, then the digital signature is correct.

***Proof:***

By the equation (22), we have

$$g_2^{r_6} \equiv g_2^{(ba+cs)r_5} \equiv g_2^{x_0 r_4 + (kb+c)s_1} \equiv g_2^{x_0 r_4} g_2^{(kb+c)s_1} \pmod{n} \quad (27)$$

There are 2 parts in last term of the equations above, consider that the first part, by equation (4) that we have

$$g_2^{x_0 r_4} \equiv (g_2^{x_0})^{r_4} \equiv y_0^{r_4} \pmod{n} \quad (28)$$

Consider that the second part,

$$g_2^{(kb+c)s_1} \equiv g_2^{(kb)s_1} g_2^{cs_1} \equiv \left((g_2^k)^b\right)^{s_1} g_2^{cs_1} \quad (29)$$

$\equiv ((r_1)^b)^{s_1} g_2^{cs_1}$, by the equation (5).
$\equiv r_1^{bs_1} (g_2^c)^{s_1} \equiv r_1^{bs_1} r_5^{s_1}$, by the equation (17).
$\equiv r_3^{s_1} r_5^{s_1} \equiv (r_3 r_5)^{s_1}$, by the equation (8).
$\equiv r_4^{s_1} \pmod{n}$, by the equation (18).

Combine these equations (28) and (29), we obtain

$$g_2^{r_6} \equiv y_0^{r_4} r_4^{s_1} \pmod{n}$$

and hence,

$$g_2^{m+r_6} = g_2^{cE} E^{s_2} \pmod{p_0}$$

Therefore, both equations (25) and (26) are valid.

When we want to check if the message $m$ has been sent by $u_i$ or not. It needs some parameters. And hence we obtain the following Lemma.

**Lemma 2** Assumes that $u_0, u_i$ are honest, only $u_0$ can prove the message $m$ was sent from $u_i$ by the equation (8).

***Proof:***

Note that,
(a) $r_1^b \equiv r_3 \pmod{p_0}$ from the equation (8)
(b) $u_i \to R : \{m, c, E, r_4, r_6, s_1, s_2\}$ from the equation (24).
(c) $u_0 \to u_i : a, s$ from the equation (12).
(d) $r_4 \equiv r_3 r_5 \pmod{p_0}$ from the equation (18).
(e) $ba \equiv x_0 r_3 + (kb)s \pmod{n}$ from the equation (15).
(f) $g_2^k \equiv r_1 \pmod{p_0}$ form the equation (5).
(g) $g_2^{r_6} \equiv y_0^{r_4} r_4^{s_1} \pmod{n}$ from the equation (25).
(h) $g_2^{m+r_6} = g_2^{cE} E^{s_2} \pmod{p_0}$ from the equation (26).

Consider that the equation (19) $s_1 \equiv r_5 s \pmod{n}$, we knew $s_1, s$ because of the equations (24) and (12). Hence, we can get $r_5$ and hence we can get $r_3$ by means of the equations (24) and (18). Finally, we must compute $b$ (just $u_i$ knew this parameter) because $u_0$ has no idea about $b$.

Consider that the equation (15), $a, x_0, r_3, k$ and $s$ is known. It is not easy to get $b$ by other people, except the manager of group $u_0$. Actually, it is a Discrete Logarithm Problem (DLP), when someone just knows $r_1, r_3$ by the equation (8).

We conclude that $u_0$ can get $b$, because he has known already a part of parameters from $u_i$ and he has his own parameter $k$. Therefore, after check the equations (25) and (26), we can to say, the message is send by $u_i$ exactly.

**Lemma 3** An attacker intercepts the message passed by the digital signature adapting the method of Willy Susilo. It will not succeed.

***Proof:***

Note that,
(a) $u_i \to R : \{m, c, E, r_4, r_6, s_1, s_2\}$ from the equation (24).
(b) $g_2^{r_6} \equiv y_0^{r_4} r_4^{s_1} \pmod{n}$ from the equation (25).
(c) $g_2^{m+r_6} = g_2^{cE} E^{s_2} \pmod{p_0}$ from the equation (26).

Assume that an attacker **A** intercepts the message as the equation (24) shows.

Due to **A** having no idea about the parameters $x_0, r_4$, then we assume,

$$g_2^{x_0'} \equiv y_0 \pmod{p_0} \tag{30}$$

$$g_2^{r_4'} \equiv r_4 \pmod{p_0} \tag{31}$$

**A** can easily forge $m^*$ for suitable parameters $\{c', E', s_2'\}$ such that both equations (25) and (26) are valid. In other words, the digital signature passes the test of Lemma 2. After the procedure of Lemma 2 and hence, a non-trivial factor of $n$ can be found by computing $GCD(b, b^*, n)$. We note that the probability of $b$ being equal to $b^*$ is $1/q_0$. Therefore, it is proved that the $m^*$ is not sent by the member of group.

## 5  Conclusion and future work

In this paper, we propose a novel FSGS scheme. This scheme integrates the features of two types of digital signatures, which strengthens its security level under the group-signature system. It ensures that group members can prove that a digital signature is indeed a forgery after supercomputer forgery attacks. In addition to the technology of integrating two digital signatures, this work also contributes three proposed lemmas and proves that they are indeed feasible. Lemma 1 is a method to verify a FSGSS digital signature. Lemma 2 is used by the group manager, when needed, to determine the identity of the group member originally creating a digital signature. Finally, this paper proposes Lemma 3. When the digital signature is found to be forged, members of the group can prove this fact.

In addition, the ultimate goal of the FSGSS is to stop using the same key immediately after the discovery of a forgery attack to avoid the same attack happening again. That is, the "key" used in this paper is the parameter $b$ used by $u_i$. If we need to change the parameters each time after an attack, the process of replacing the parameters is equivalent to re-executing the exchange parameter program. Therefore, in future work, if we cannot directly expose the key $b$, we could still prove that a certain number of signatures are forged, which would enhance the efficiency of the GFSS scheme.